# On Simon Mayr's (Marius) alleged discovery of Jupiter's satellites

## Gabriele Vanin
*Independent scholar (rheticus@tiscali.it)*

**Abstract**
In 1614, the German astronomer Simon Mayr published his claim to have discovered the satellites of Jupiter. Writing in the treatise *Mundus Jovialis*, Mayr made his assertion in a convoluted but unequivocal manner, earning the displeasure of Galileo Galilei, who published his harsh protest in 1623 in *Il Saggiatore*. Though Galileo's objections were fallacious in some respects, and though numerous scholars took to the field to prove Mayr's claim, none ever really succeeded, and the historical evidence remains to Mayr's detriment. On the basis of such historical evidence, including comparisons between *Mundus Jovialis* and Mayr's earlier works, Mayr's independent discovery of the satellites can be ruled out. Indeed, it is very likely that he never observed them before 30 December 1610, nearly a year after Galileo. The absence of a corpus of Mayr's observations and the inaccuracy of his tables are also puzzling.

**Introduction**

In 1623, in *Il Saggiatore*,[1] Galileo harshly attacked the German astronomer Simon Mayr, accusing him of plagiarism, because Mayr had claimed the discovery of Jupiter's satellites in his *Mundus Jovialis*, published in 1614.[2] No one took Mayr's claim very seriously from the outset, either because of the timing of the claim, which was very late, or because *Mundus Jovialis* was not widely circulated and probably also because Galileo's defence appeared convincing.

**Galileo's reaction**

In order to demonstrate the inconsistency of Mayr's claims, Galileo focused on the sixth phenomenon described by Mayr[3]—that is, Mayr's assertion that the satellites were in a line parallel to the ecliptic (and therefore appeared aligned with Jupiter in a straight line) only when they were at their maximum elongations. Outside of these, they appeared to be shifted northwards in the lower parts of their orbits, and southwards in the upper parts,[4] due to the fact that their orbits are inclined southwards in the upper parts (i.e. furthest from the Sun), and northwards in the lower parts (closest).[5] In other words, according to Mayr, the planes of the orbits were inclined along a line parallel to the Jupiter-Sun direction.

---

[1] Rome: Mascardi, 1623.
[2] Simon Mayr, *Mundus Jovialis* (Nuremberg: Lauer, 1614).
[3] Ibid., C4rv.

     Galileo objected that the orbits were parallel to the ecliptic, not inclined, and, consequently, that the satellites could appear in a straight line at any distance from Jupiter. Moreover, at certain times they appeared shifted in a direction opposite to Mayr's assertion (i.e., southwards in the lower parts of the orbits and northwards in the upper ones). In other words, if Jupiter were on the ecliptic, the satellites would always appear to be arranged in a straight line. If Jupiter were above the ecliptic, the satellites would always appear to be positioned northwards in the lower parts and southwards in the upper parts, and, if Jupiter were below the ecliptic, the satellites would always appear to be positioned southwards in the lower parts and northwards in the upper parts.

     In fact, Jupiter was below the ecliptic when Galileo began to observe it, and its satellites should therefore have appeared opposite to Mayr's claim. Because Jupiter remained so southern for more than two years, Galileo concluded, it meant that Mayr must not have observed Jupiter for the first time until more than two years after him. Not only that, Galileo continued, but the fact that Mayr said that he had never observed the satellites in a straight line except at their maximum elongations meant that he had never observed them at all, because, for example, they were continuously in this position at least from mid-February to mid-June 1611. Finally, he accused Mayr of misleading readers by not warning them that he had used the Julian calendar and, therefore, that even Mayr's first observation could not be considered earlier than Galileo's.

     In reality, the orbits of the satellites are not parallel to the ecliptic, but are actually inclined with respect to Jupiter's orbit around the Sun. However, this inclination is certainly not always directed toward the Sun, as Mayr thought, but is fixed in space. The different latitude of the satellites (i.e., the deviation from the line passing through Jupiter's centre) is therefore the result both of this inclination and of Jupiter's ecliptic latitude. Therefore, both Galileo's and Mayr's hypotheses were wrong. Moreover, it is not true, as Galileo claimed, that Jupiter remained under the ecliptic for the first two years of observation but only for the first year. Here are the presentations of the satellites from the end of 1609 to 1614:[6]

     1) 1609-10: Jupiter always under the ecliptic, satellites to the south in the lower parts of the orbits, to the north in the upper ones.

     2) 1610-11: Jupiter always above the ecliptic, satellites southward in the lower parts, northward in the upper parts.

     3) 1611-12: Jupiter always above the ecliptic, satellites southward in the upper parts, northward in the lower parts.

---

[4] *Il Saggiatore*, pp. 3-6.
[5] Because Mayr's words were not clear, it was actually Galileo who deduced the cause; the concept was hardly better expressed in E3v.
[6] Simulated using *Starry Night Pro Plus 8* software. The measurements were made for the first quadrature, the opposition, and the second quadrature, thus covering the entire duration of each presentation.

4) 1612-13: Jupiter always above the ecliptic, satellites to the north in the lower parts, to the south in the upper parts.

5) 1614: Jupiter always above the ecliptic, satellites to the north in the lower parts, to the south in the upper parts.

As can be seen, therefore, the trend was as predicted by Galileo's hypothesis in Cases 1, 3, 4, and 5, but not in Case 2. Galileo's prediction of the presentation of the orbits with Jupiter on the ecliptic was also wrong. In August 1610, for example, the satellites appeared southward in the lower parts of their orbits and, in November 1616, northward in the lower parts.

It is nonetheless true that in the first two presentations the satellites appeared southward in the lower parts of their orbits and northward in the upper parts, even if not for the reasons assumed by Galileo. This is a strong indication that Mayr did not observe the satellites in the first two presentations or, at least, that he did not observe them as closely as he wrote in *Mundus Jovialis*.

On the other hand, Galileo's remark about the straight-line presentation of the satellites for four consecutive months in 1611 is wrong. Even Galileo, as also Bosscha pointed out,[7] reported latitudinal deviations of the satellites, in the form of drawings or notes in his observation journal, on at least eight occasions: 7-11-16-18-26 March, 17 April, and 6 and 29 May.[8]

Instead, Galileo was right in his observation regarding the calendar difference, although the issue, as we shall see, is more complex.

To many, Galileo's reaction and tone appeared out of proportion. His great disappointment enters into the account, however, because he was convinced Mayr had plagiarised another of his discoveries, the proportional compass, in a 1607 work,[9] which, as historiographic research[10] and the sentence of the court in Padua[11] demonstrated, turned out to be a clumsy remake of Galileo's *Il compasso geometrico e militare*.

Baldassarre Capra, Mayr's pupil, appeared to be the author of the work, but he dedicated the book to Mayr's patron, Margrave Joachim Ernest and, during the evidentiary investigation before the Court of Padua, displayed complete ignorance of the most basic principles of geometry, making clear that he did not even know what his treatise was about.[12] It was evident that Capra could not be the author but rather that the treatise had been written by Capra's mathematics teacher, Mayr, whom Capra frequented assiduously during the German's years in Padua, from December 1601 to July 1605.[13]

---

[7] Johannes Bosscha, "Réhabilitation d'un astronome aalomnié," *Archives Néerlandaises des Sciences Exactes et Naturelles*, 12 (1907) 258-307, 490-528, p. 259.

[8] Galileo Galilei, *Opere*. Edited by Antonio Favaro. (Florence: Barbera, 1890-1909), 3.2, pp. 441-5.

[9] *Usus fabrica circini cuiusdam proportioni*, in Galilei, *Opere*, 2, pp. 425-511.

[10] Gio. Battista Clemente de' Nelli, *Vita e commercio letterario di Galileo Galilei* (Lausanne: 1793), p. 113. Antonio Favaro, *Galileo Galilei e lo Studio di Padova* (Florence: successori le Monnier, 1883), 1, pp. 212-48. Adolf Müller, *Galileo Galilei: studio storico-scientifico* (Rome: Bretschneider, 1911), pp. 39-42. Emil Wohlwill, *Galilei und sein Kampf für die copernicanische Lehre*, 2 (Leipzig: Voss, 1926), pp. 416-426. Ludovico Geymonat, *Galileo Galilei* (Turin: Einaudi, 1972), 6th ed., pp. 38-39. Stillman Drake, *Galileo: una biografia scientifica* (Bologna: Il Mulino, 1988), p. 183. Eileen Reeves, *Galileo's glassworks* (Cambridge-London: Cambridge University Press, 2008), pp. 101-2.

[11] Antonio Favaro, *op. cit.* (note 10), 2, pp. 274-5.

[12] Galileo Galilei, *Difesa contro le calunnie e le imposture di Baldessar Capra*, in *Opere*, 2, pp. 551-559.

According to Galileo, Capra also directly pointed to Mayr as the real author,[14] both during the deposition and afterwards. Galileo's wrath was increased by the fact that Capra-Mayr's work sought to make Galileo out to be a plagiarist, in front of his city, his friends, his colleagues, and his students. It was also not the first time Capra had clashed with Galileo. In 1604, on the occasion of the appearance of the famous supernova of that year, Capra and Mayr had been the first to observe it from Padua. Galileo had made it the subject of three packed lectures at the university, in which he had correctly mentioned the two men's discovery.[15] But Capra and Mayr were evidently not satisfied. Perhaps envious of the success of the lectures, they published a treatise on the subject[16] containing outrageous comments about Galileo. Probably because of Capra's youth, Galileo did not respond to the attack. Although only Capra was once again credited for the work, it was Mayr himself who later revealed that he was the author, as Klug has shown:[17]

> ... In Padua, Italy, I also dictated to my disciple in mathematics, Balthasar Capra, a nobleman from Milan, a manuscript treatise, which he translated in his own name, for my benefit, because there a distinguished Professor of Philosophy had published senseless things against the observations of astronomers.[18]

Galileo remained silent for a long time after the publication of *Mundus Jovialis*, but, when he finally had the opportunity,[19] he decided to launch an attack.

**In defence of Mayr**

From the eighteenth century onwards, however, Mayr found an increasing number of champions in Germany.[20] The best known of these was probably Johann Gabriel Doppelmayr. Alexander von Humboldt openly supported Mayr in his *Kosmos*,[21] moreover, and von Humboldt's great influence on German culture meant that, from that time on, Mayr's name was increasingly referred to as the true discoverer of Jupiter's satellites in secondary sources and popular treatises.

---

[13] 13. Hans Gaab, "Concerning the biography of Simon Marius" in Hans Gaab and Pierre Leich (eds.), *Simon Marius and his research.* (Cham: Springer, 2018), pp. 88 and 104.

[14] Galilei, *op. cit.* (note 12), pp. 540, 554, 594. On the other hand, this can also be deduced from the dedication of the *Usus fabrica* (pp. 429-31).

[15] Ibid., p. 520.

[16] Baldesar Capra, *Consideratione astronomica circa la stella nova dell'anno 1604*, in Galilei, *Opere*, 2, 285-305.

[17] Josef Klug, "Simon Mayr aus Gunzenhausen und Galileo Galilei," *Abhandlungen der mathematisch-physikalischen Klasse der Königlich Bayerischen Akademie der Wissenschaften*, 22 (1906), 385-526, pp. 407-8.

[18] Simon Mayr, *Prognosticon astrologicum 1623* (Nuremberg: Lauer), A2v.

[19] After 1613, Galileo published nothing more, partly because of the edict of 1616 that forbade him from openly professing Copernicanism. *Il Saggiatore*, already completed in November 1621, was not published until November 1623.

[20] Klug, *op. cit.* (note 17), p. 430.

[21] Alexander von Humboldt, *Kosmos* (Stuttgart and Tübingen: Cotta, 1847 and 1850), 2, pp. 356-7, 509-10; 3, pp. 315, 354-5, 522.

The first person in Germany to seriously oppose this trend was Christian Frisch,[22] the editor of Kepler's *Opera Omnia.* Among other things, Frisch noted that Mayr had mentioned the satellites in his dedication (dated 30 June 1612) to the 1613 *Prognosticon* but did not claim their independent discovery:

> I have also mentioned in *Prognostico*[23] at various times the 4 New Jovian Planets, together with their general hypotheses, and that the period of the fourth or outer was investigated and tabulated by me,…[24]

A notable turning point in the controversy came in 1900, when the Société Hollandaise des Sciences proposed a competition to publish the best paper proving that there had been no plagiarism by Mayr. The only entry, however, written by Josef Klug of Nuremberg, went in completely the opposite direction. The jury decided not to publish it because of its length of over 200 pages, its excessively biased tone, and its inadequate adherence to scientific standards. What is more, two of the organisers of the competition, Johannes Bosscha and Jean Abraham Chrétien Oudemans, decided to take it upon themselves to write an article for Mayr's rehabilitation, which was published in 1903.[25] In it, Bosscha and Oudemans sought to demonstrate that Mayr did not attempt to appropriate Galileo's discovery but only "… to assert in good faith that one of these wonders, the system of the four satellites of Jupiter, was seen and recognised by him almost at the same time as Galileo."[26] Consequently, they argued, Mayr should be absolved of Galileo's disdainful accusations. However, they chose to diminish Galileo's merits in order to increase Mayr's.

I would like to highlight just a few of Bosscha's and Oudemans' most surprising statements, in which they attempted to demonstrate the inferiority of Galileo's telescopes compared to Mayr's: 1) Galileo did not make any improvements to the telescope, which remained the same since the Dutch invention; 2) he did not build his own telescopes;[27] 3) his amplification of magnification was not an improvement in the power of the instrument; 4) even with current technology no optician could build a Dutch telescope of a 30x magnification because it would be unserviceable; and lunar drawings prove the poor quality of the lenses used by Galileo.[28]

Still, Bosscha and Oudemans brought no new documents to the attention of scholars in favour of Mayr, beyond its own statements in the *Mundus Jovialis*. As the only evidence in favour of Mayr's independent discovery, they presented the testimony of his patron, General Hans Philip Fuchs von Bimbach, and, by implication, of the Margraves of Brandenburg, to

---

[22] *Johannis Kepleri astronomi opera omnia* (Frankfurt and Erlangen: Heider & Zimmer, 1859), 2, p. 470.
[23] He refers to *Prognosticon* for 1612, published the previous year.
[24] *Prognosticon astrologicum 1613* (Ansbach: Lauer, 1613), A2v. In this *Prognosticon* Mayr also reports for the first time the periods of the four satellites (A4r).
[25] Jean Abraham Chrétien Oudemans and Johannes Bosscha, "Galilee et Mayr," *Archives Néerlandaises des Sciences Exactes et Naturelles*, 2(8), (1903), 115-89.
[26] Ibid., p. 117.
[27] Ibid., p. 130.
[28] Ibid., p. 131. With regard to all of these aspects, allow me to refer to my article, "Much better than thought: observing with Galileo's telescopes." *Journal for the history of astronomy*, 46(4), (2015), 441-68, from which the primary sources, which are not always easy to identify, can be readily traced.

whom *Mundus Jovialis* was dedicated, and to whom Fuchs von Bimbach was the most influential advisor.[29]

The decision of the Société Hollandaise des Sciences jury caused Antonio Favaro, the editor of the National Edition of Galileo's *Opere*, to express his regret, in the spirit of freedom of research, that he could not read Klug's arguments.[30] However, Klug was able to publish his research, revised and corrected, in 1906.[31] In contrast to Bosscha and Oudemans, Klug presented several new documents, which included unknown writings by Mayr and others and letters from himself and others, and made a thorough analysis of Mayr's entire scientific work.

On some points Klug handled his compatriot's contributions with fundamental fairness, but, with regard to others, he was certainly too critical. He stigmatised Mayr's mystical attitude and belief in astrology, for example,[32] traits that were known among other scholars of the time, including in Kepler himself. In other instances, he was unfair or did not assess Mayr's work correctly (e.g., failing to recognise the correctness of Mayr's observations regarding Mercury[33] or comparing the dimensions Mayr gave for Jupiter with modern values).[34]

He, too, was on the whole overly biased, this time in favour of Galileo, but he analysed very seriously the many inconsistencies and contradictions in Mayr's writings and, among other things, clarified the relationship between Mayr and Kepler and carried out a thorough examination of the technical level of the telescopes of the time.[35]

In the strongest part of Klug's article, he tried to show that all the progress Mayr claimed in the study of satellites was actually taken from Galileo's publications:[36] the publication of *Sidereus Nuncius* in March 1610 was followed by *Prognosticon* for 1612 with Mayr's first report of satellites; the publication of the periods in the *Discorso delle cose che stanno in su l'acqua* in April 1612 was followed by Mayr's first publication of the periods in *Prognosticon* for 1613; the "costituzioni delle Medicee," or charts presenting the position of the satellites from 1 March to 8 May 1613, published in Galileo's *Lettere Solari*, were followed in 1614 by *Mundus Jovialis*, with the publication of the tables of the satellites. However, although ingenious and based on numerous clues, Klug's reconstruction cannot be confirmed.

The publication of Klug's article provoked a further outcry in Dutch circles, and Bosscha published a further contribution with an even stronger stand in favour of Mayr.[37] In it, he classified Klug's reconstruction as generally inconsistent, tendentious, and imaginary. Although he succeeded in proving Klug wrong on some points, he ignored other questions, especially those concerning the serious contradictions among historical documents. He also failed to effectively dismantle Klug's reconstruction of Mayr's dependence on all of Galileo's published sources. He also misinterpreted some of Klug's observations. For example, Klug stated on p. 439 of his article that "There is not even a single testimony that Marius saw the satellites at that time," and it is clear from the context that he speaks of the end of 1609, when Mayr says he first observed Jupiter. Bosscha, however, inexplicably added "1614" in brackets to Klug's verbatim quotation.[38]

---

[29] Wolfgang R. Dick, "Hans Philip Fuchs von Bimbach (ca. 1567-1526), patron of Simon Marius," in Hans Gaab and Pierre Leich (eds.), *op. cit.* (note 13), 139-77, p. 147.
[30] Antonio Favaro, "Galileo Galilei and Simone Mayr," *Bibliotheca Mathematica*, *3*(2) (1901), 220-223.
[31] Klug, *op. cit.* (note 17).
[32] Ibid., pp. 400-1.
[33] Ibid., p. 475-6. Demonstrated instead by Bosscha, *op. cit.* (note 7), pp. 518-27.
[34] Ibid., p. 477.
[35] Ibid., pp. 455-62.
[36] Ibid., pp. 443-8, 462-71.
[37] Bosscha, *op. cit.* (note 7).

In the footnote at the end of p. 471, Klug quoted Mayr's statement in the second edition of *Mundus Jovialis* in which Mayr denied having read any of Galileo's works other than *Sidereus Nuncius.* Bosscha then pedantically accused Klug[39] of distorting Mayr's thought because Mayr had actually used a stronger expression ("I swear that I have not read") rather than "I deny that I have read." In fact, however, Bosscha pretended not to understand that Klug disbelieved Mayr's statement anyway given that Klug had presented a great deal of evidence to the contrary. In this context, the fact that Klug had used a milder expression perhaps acted in Mayr's favour.

Bosscha stated that Klug believed that Mayr had used elongations recorded only from one revolution to the next for the calculation of the periods.[40] Conversely, on p. 103 of his treatise, Klug stated that "…even a comparison of observations which are very far apart could not lead to the desired objective."

Above all, however, Bosscha continued to try to discredit Galileo, denigrating the quality of his telescopes, proposing confused theories and incorrect formulas on the magnitude of the field of the Galilean telescope,[41] and misinterpreting some of Galileo's observations, such as when, by mistranslating his notes on observations of the Medicean stars of 15 January 1613, he attributed to Galileo an estimate of 40" of the apparent diameter of a satellite, when he was obviously referring to the separation between two satellites.[42]

Probably also because of its didactic title, Bosscha's article caused a certain resonance in the United Kingdom where several authors, who had read Oudemanns' and Bosscha's works but evidently not Klug's, sided with Mayr.[43]

Although a long and articulate critical contribution on the question by Emil Wohlwill was published posthumously in 1925 as an appendix to the second volume of his work on Galilei,[44] scholars on the other side of the Atlantic later embraced Mayr's cause,[45] and the hitherto little-used names he had proposed for the Jovian satellites, Io, Europa, Ganymede, and Callisto,[46] began to spread.[47] Some popular texts even began to place Mayr's name alongside Galileo's as the satellites' discoverer.[48]

---

[38] Ibid., p. 276.
[39] Ibid., p. 281.
[40] Ibid., p. 269.
[41] Ibid., pp. 290, 291 and 501. In particular, he states that the field is not proportional to the aperture of the telescope and that Galileo used diaphragms in front of the objectives, not to improve the images, but to reduce the field! (p. 291).
[42] Ibid., p. 287. Galilei, *Opere*, 3.2, p. 428.
[43] William Thynne Lynn, "Simon Mayr and the satellites of Jupiter," *The Observatory*, 26 (1903), 254-6; *idem*, "Galileo and Mayr," *The Observatory*, 27 (1904), 63-4; *idem*, "Simon Mayr," *The Observatory*, 32 (1909), 355-6; Arthur Octavius Prickard, "Note on "Simon Mayr" and the 'Mundus Jovialis'*,"* *The Observatory*, 40 (1917), 119-22; Joseph Horsfall Johnson, "The discovery of the first four satellites of Jupiter," *Journal of the British Astronomical Association*, 41 (1931), 164-71; Evelyn G. Hogg, "The moons of Jupiter: Mayr and Galileo," *Journal of the Royal Astronomical Society of Canada*, 25 (1931), 6-9.
[44] Emil Wohlwill, *op. cit.* (note 10), 2, 343-426.
[45] Evelyn G. Hogg, "The moons of Jupiter: Mayr and Galileo," *Journal of the Royal Astronomical Society of Canada*, 25 (1931), 6-9; Samuel G. Barton, "The names of the satellites," *Popular Astronomy*, 54 (1946), 122-30; *idem*, "Discovery and naming of Jupiter's satellites," *Astronomical Society of the Pacific Leaflets*, 5(214) (1946), 111-8.

The most recent episode in this wave of revisionism was the birth of a vast opinion movement in Nuremberg, close to Mayr's birthplace of Gunzenhausen, on the occasion of the fourth centenary in 2014 of the publication of *Mundus Jovialis*, which led to a wide-ranging project for the definitive rehabilitation of Mayr, including the creation of a website in various languages and the celebration in 2018 of a conference and the publication of the proceedings.[49]

**The example of sunspots**

Apparently, Mayr's claim, made a full four years after the first discovery and after numerous independent confirmations by other astronomers, should not even have been considered by the scientific community. This applies today but was also the case four centuries ago. An example of this is the Galileo and Christoph Scheiner "sunspot affair" in which the precedence of the one over the other was a matter of no more than a few months or even a few weeks. Here, too, a heated controversy arose. Scheiner, in his first letter to Markus Welser on 12 November 1611, said that he had observed sunspots for the first time with a friend (Johann Baptist Cysat) about seven or eight months earlier, therefore in March-April 1611.[50] Scheiner's letters were published on 5 January 1612. Galileo's account, in contrast, contains great contradictions. In the *Prima lettera sulle macchie solari* dated 4 May 1612, he stated that he had observed them for the first time "18 months ago,"[51] (i.e., in November 1610), and he repeated the figure in a 2 June 1612 letter to Maffeo Barberini.[52] In the *Dialogo*, on the other hand, Galileo dated the discovery to before September 1610, when he was still in Padua,[53] a date that is consistent with a 27 September 1631 letter written to him by Father Fulgenzio Micanzio.[54] This evidence, however—a comment on a subject that had already given rise to much controversy and coming as it did from a trusted friend nearly twenty years after the fact—cannot be considered reliable. Among other things, the possibility exists that Galileo had requested Micanzio's testimony to reinforce the date Galileo had given in the *Dialogo*.

The testimony of Galileo's disciple, Viviani, is even less reliable. Viviani placed the observations in the same period but wrote in 1654, citing many authoritative witnesses who were, alas, all close friends of Galileo.[55] Viviani also reported that, at the end of March 1611, Galileo had shown the sunspots to many prelates and cardinals during his visit to Rome,[56] something mentioned by Galileo himself in his *Prima lettera sulle macchie solari*.[57] This is partially confirmed by three letters written to Galileo by Gio Battista Agucchi and Lodovico Cigoli, but more than a year later, and where it is said that the spots were not shown, but it was talked about them.[58]

---

[46] Mayr, *op. cit.* (note 2), B2.
[47] Probably mainly because of Barton's articles, in the first of which he stated "Those who favored Galileo and who thought Marius an imposter declined to use the latter's names lest their use be regarded as a recognition of his claims,…" (p. 125) and, in the second, "…the taboo against the names assigned by Marius has been removed…" (p. 118).
[48] Examples include Patrick Moore, *The Guinness book of astronomy* (London: Guinness Publishing, 1979); J. Kelly Beatty and Andrew Chaikin (eds.), *The new solar system* (Cambridge: Sky Publishing Corporation, 1981); and John Gribbin, *Companion to the cosmos* (London: Weidenfeld & Nicolson, 1996).
[49] Gaab and Leich (eds.), *op. cit.* (note 13).
[50] *Tres epistolae de maculis solaribus* (Augsburg: 1612), B2r.
[51] Galileo Galilei, *Istoria e dimostrazioni attorno alle macchie solari* (Rome: Mascardi, 1613), p. 11.
[52] Galilei, *Opere*, 11, p. 305.
[53] *Dialogo di Galileo Galilei Linceo* (Florence: Landini, 1632), p. 337.
[54] Galilei, *Opere*, 14, p. 299.

      The Agucchi and Cigoli accounts, however, would contradict Cardinal Bellarmino's famous 16 April 1611 letter in which he asked the mathematicians of the Collegio Romano for confirmation of Galileo's new discoveries,[59] among which the question of the sunspots was not mentioned. Perhaps the surest attestation in favour of Galileo, because it was not produced *a posteriori*, is contained in the *Discorso delle cose che stanno in su l'acqua*,[60] published in April 1612 but probably finished in October of the previous year.[61] Thus Scheiner probably preceded Galileo. In any case, the only reason for which reliable evidence exists—as opposed to the more-or-less reliable witness accounts of friends and acquaintances—is that the observations were followed fairly quickly by publication.

      Scheiner, however, may have been preceded by Johannes Fabricius, who first saw spots on the Sun in the spring or perhaps late winter of 1611 and was the first to publish the news.[62] But all were certainly preceded by Thomas Harriot, who observed and drew the Sun with three large spots in his observation logbook on 8 December 1610 (Julian date).[63]

**The attempted theft**

There is no doubt that Doppelmayr, other eighteenth-century German scholars, and von Humboldt sided with Mayr on the basis of their reading of *Mundus Jovialis* and trusted what he had stated in it. First of all, the title clearly indicated that the discovery of the satellites occurred in 1609: *The world of Jupiter discovered in the year 1609 by means of a Belgian telescope*. Three equally explicit statements appeared inside:

> … towards the end of November… for the first time I looked at Jupiter… and saw small stars which sometimes followed, sometimes preceded Jupiter in a straight line with it.[64]

---

[55] *Racconto istorico di Vincenzio Viviani*, in Galilei, *Opere*, 19, p. 611.
[56] Galilei, *Opere*, 11, p. 305.
[57] Galilei, *op. cit.* (note 51), p. 11.
[58] Galilei, *Opere*, 11, pp. 329, 418, and 424.
[59] Ibid, pp. 87-88.
[60] Galilei, *Opere*, 4, p. 64.
[61] Antonio Favaro, *Miscellanea galileiana inedita* (Venice: Antonelli, 1887), p. 37.
[62] *De maculis in Sole observatis* (Wittenberg: Seuberlichius, 1611). The dedication of the work is dated 13 June 1611.
[63] Thomas Harriot, *Harriot Papers*, Vol. VIII: *Spots on the Sun* [HMC 241 VIII f. 1]. For reasons that are unclear, Harriot published almost none of his countless works in mathematics, physics, and astronomy. For a possible interpretation, See Allan Chapman, "The Astronomical Work of Thomas Harriot," *Quarterly Journal of the Royal Astronomical Society*, 36 (1995), pp. 104-5.
[64] Mayr, *op. cit.* (note 2), *Praefatio ad candidum lectorem*, 2v.

> ... my first observations, made in the autumn of 1609,...[65]

> ... and I still saw these stars that accompanied it throughout the month of December,...[66]

Doppelmayr, other eighteenth-century scholars, and von Humboldt certainly believed Mayr's precedence over Galileo who, as we know, first observed the satellites on 7 January 1610.[67] Indeed, in confirmation of the above, Mayr claimed to have observed the satellites almost at the same time as, or shortly before, Galileo:

> ... these stars were not shown to me by any mortal in any way, but were discovered and observed by me, by my own investigations, in Germany, almost at the same time, or a little earlier, that Galileo first saw them in Italy.[68]

In short, Galileo was only the first discoverer in Italy. Mayr continued:

> In recounting all this, I do not wish to be understood as wishing to diminish Galileo's reputation, or to wrest from him the discovery of the Jovian stars among his countrymen in Italy, quite the contrary.[69]

> The credit, therefore, for the first discovery of these stars in Italy is deservedly given to Galileo....[70]

Shortly thereafter, Mayr even thanked Galileo for publishing *Sidereus Nuncius*, which confirmed his observations![71]

It is surprising, therefore, that Oudemans and Bosscha excluded the possibility that Mayr had engaged in plagiarism. Perhaps, though, one should not speak of plagiarism in this case, although Frisch and Favaro[72] have pointed out that parts of the text of *Mundus Jovialis* seem to have been taken verbatim from *Sidereus Nuncius*. Perhaps the words "appropriation" or "misappropriation" would be more appropriate. However, the attempt at some kind of "appropriation" is also clear from the date Mayr used for his first official recording of an observation in *Mundus Jovialis*, 29 December 1609, in a single form and not double as was customary at a time when Protestant countries had not yet adopted the Gregorian calendar. It should be noted above all, as Mayr clearly states, that this observation were made after he recognized the stars as satellites of Jupiter:

> ... but little by little I arrived at the following opinion, namely, that these stars were moving around Jupiter, just as the five solar planets, Mercury, Venus, Mars, Jupiter, and Saturn revolve around the Sun.[73]

---

[65] Ibid., B4r.
[66] Ibid., *Praefatio*, 3r.
[67] Galileo Galilei, *Sidereus Nuncius* (Venice: Baglioni, 1610), p. 17r.
[68] Mayr, *op. cit.* (note 2), *Praefatio*, 3v.
[69] Ibid.
[70] Ibid.
[71] Ibid.
[72] Frisch, *op. cit.* (note 22), 2, p. 471; Favaro, *op. cit.* (note 10), 1, pp. 441-3.
[73] Mayr, *op. cit.* (note 2), *Praefatio*, 3r.

This recognition therefore predates the Julian date of 29 December 1609 (i.e., the Gregorian date of 8 January 1610). In any case, 8 January 1610 is earlier than 11 January, the date on which Galileo established that the new stars revolved around Jupiter. Even in this way, then, Mayr's attempt to claim first discovery is evident.

**A look at *Mundus Jovialis***

Had they read *Mundus Jovialis* more carefully, eighteenth-century German scholars and Von Humboldt himself could perhaps have entertained some suspicions. The treatise's scientific level leaves something to be desired, even for its time, and all the more so if we take into account the fact that the satellites had been known for four years. In general, the content is meagre, the work is full of errors and contradictions, and most of the discussion is banal with many repetitions. The title indicates that the treatise was based mainly on personal observations, and Mayr repeatedly stated that he based the work on his own observations. In the whole work, however, only three quantitative measurements are mentioned. Mayr claimed that he had not been able to publish them previously, citing improbable excuses, and noted that he would publish them in a future edition.[74] He also stated that, during December 1609, he realised that the satellites were moving around Jupiter and began to record his observations.[75] But where are these records? The comparison of the publication of his work to Erasmus Reinhold's venerable *Tabulae Prutenicae*, which takes up most of the dedication, is puzzling to say the least. Mayr said that it took him a good fourteen days, from 29 December to 12 January,[76] to understand "somehow" that the satellites were not three, but four, though he had already been observing their evolutions for several weeks. Moreover, it is only after at least another 20 days of observation (between 8 February and the end of February or beginning of March) that he is able to confirm the number of satellites.[77]

Mayr mentioned "his" method for obtaining the relative distances of the satellites from Jupiter, but did not say describe it, repeating that he would publish it elsewhere.[78] The approximation he used for π, 22/7,[79] was already wrong at the third decimal place and totally inadequate for the mathematical standards of his time.[80] Mayr gave the diameter of Jupiter as 35/60 of that of the Earth.[81] Starting from the value of the apparent diameter found by him (one arcminute), however, it should have been 201/60 taking the distances of Ptolemy, 104/60 according to Copernicus, and 70/60 according to Tycho.[82] He is unclear about the value of the maximum elongation of the fourth satellite: first he says 13 minutes, then 14, then 13 or 14, then 12.5.[83] The presentation and explanation of the fifth phenomenon presented by Jupiter's satellites is not at all clear, and he loses himself in considerations not very understandable and useless for the purpose.[84]

---

[74] Ibid., *Praefatio*, 2r; F2v.
[75] Ibid, *Praefatio*, 3r.
[76] Ibid.
[77] Ibid.
[78] Ibid., 3v.
[79] Mayr, *op. cit.* (note 2), Av.
[80] One could not expect him to use Van Ceulen's to the 35th decimal place, published in 1610, or Van Ceulen's to the 20th digit, published in 1596, or Kashani's to the 17th (1424), but at least Viete's to the 9th (1593) or Ptolemy's to the 4th.
[81] Mayr, *op. cit.* (note 2), Ar.
[82] See Albert Van Helden, *Measuring the universe* (Chicago-London: University of Chicago Press, 1986), pp. 27, 46 and 50.

Mayr stated that he required seven to eight months to establish that the fourth satellite orbited in seventeen days and a year to establish that the third orbited in seven days. He then stated that, in March 1611 (i.e., just two and a half months later), he came to establish the periods 16d18h and 7d3h53m for them.[85] Not only that, but just afterwards, as Drake noted,[86] he says that he had already compiled tables of the satellites when he read the *Sidereus Nuncius*, in June 1610 indeed![87]

In order to simplify the calculations, Mayr reduced the ratio between the size of the Earth's and of Jupiter's orbits, according to Copernicus, from 11.5/60 to 11/60, but we cannot see why.[88] He claimed that he would later explain the method by which he calculated a table of equations (for what purpose is not clear), but he did not.[89] He claimed to have arrived independently at the Tychonic system—twenty-six years after Brahe published it![90]

**Mayr contradicts himself**

It can also be ruled out that Mayr could have made the discovery, if not first, completely independently of Galileo. Mayr himself is the first to deny this in his writings. In the *Prognosticon* for the Year 1612, which Frisch probably did not know[91] and whose dedication is dated 1 March 1611, Mayr mentions the satellites as "the four new planets around Jupiter,"[92] (i.e., stars whose existence was is already known). He did not claim any discovery, but wrote that:

> … the planet Jupiter with its planets, which have their course around it, but are so small that they cannot be seen without the Dutch instrument, as I have observed and seen them many times with such an instrument, from the end of December 1609 until 1610, about which I have also written to Mr. David Fabricius, in East Frisia, and to Mr. M. Odontius in Altdorf.[93]

And further on:

---

[83] Mayr, *op. cit.* (note 2), Av, Br and Cr.
[84] Ibid., B3v, C2v, C3r and C3v.
[85] Ibid., C2r.
[86] Stillman Drake, "Galileo and satellite prediction," *Journal for the history of astronomy*, 10 (1979), p. 86.
[87] Mayr, *op. cit.* (note 2), C2v.
[88] Ibid., E3r.
[89] Ibid., C3r.
[90] Ibid. The same system had been proposed by the two German astronomers Helisaeus Roeslin and Nicolai Ryemers Bär, at the same time as Tycho. See J.L.E. Dreyer, *Tycho Brahe* (Edinburgh: Black, 1890), pp. 184 and 274.
[91] Klug wrote that a copy was only found in 1902 in the Nuremberg library, *op. cit.* (note 17), p. 518, note 1.
[92] *Prognosticon astrologicum 1612* (Nuremberg: Lauer, 1612), A3r.
[93] Ibid., B1v-B2r.

Such a thing has never been observed since the beginning of the world. Galileo Galilei, a mathematician from Padua, had already published a treatise on these new planets, as was also indicated at the beginning of the *Practica*...[94]

As can be seen, Mayr here postpones his first sight of the satellites from the end of November, as reported in *Mundus Jovialis*, to the end of December, and he makes it coincide with the beginning of his telescopic observations;[95] in *Mundus Jovialis*, in contrast, he indicates that these observations began in the summer of 1609.[96] Moreover, no trace of the letters addressed to Fabricius and Odontius has been found, although Odontius mentions a letter from Mayr in a letter to Kepler dated 24 November 1611,[97] in which he reports an observation of the four satellites made during the lunar eclipse of 19-20 December 1610 (29-30 December in Gregorian date) by the man he called *non inceleber mathematicus.* As Klug has shown,[98] a fragment of another letter from Mayr to Nikolaus Vicke, attaché at the imperial court in Prague, was reported in a letter Vicke wrote to Kepler on 6 July 1611. In it, Mayr says:

> ... I give an account of the four new Jovian planets, which revolve around Jupiter, like the other planets with the Sun, but at unequal distances and periods. I have already found the periods of the two outermost ones, and constructed tables, so that from now on it will be very easy to know at any time how many minutes they are to the right or left of Jupiter.[99]

In short, even in these letters Mayr makes no claim to discovery. As Wohlwill has shown,[100] Fabricius recalled Galileo's discovery of Jupiter's satellites in the *Prognosticon* for the year 1615 (the dedication is dated June 1, 1614) and says that Mayr was also studying them and hoped soon to communicate his results.[101] This makes clear that Fabricius knew Mayr was preparing a publication on the Jovian planets but that he knew nothing about Mayr's independent claim to have discovered them.

There is absolutely nothing about Jupiter's satellites in Mayr's prognostics of previous years. Nothing in 1609, when Mayr could at least have reported his fanciful account of learning, in autumn 1608, of a new instrument invented in the Netherlands, which he included in *Mundus Jovialis*. And nothing especially in 1610 and 1611. These prognostics had not been found at the time of Bosscha, Klug, and Oudemans but were later discovered by Ernst Zinner.[102] Their dedications bear the dates 31 March 1608, 13 January 1609, and 13

---

[94] Ibid., C3v.
[95] Ibid., A2v.
[96] Mayr, *op. cit.* (note 2), *Praefatio*, 2v.
[97] Michael Gottlieb Hanschius (ed.), *Johannis Keppleri aliorumque epistolae mutuae* (Leipzig: 1718), p. 304.
[98] *Op. cit.* (note 17), p. 419.
[99] Hanschius, *op. cit.* (note 97), pp. 326-7.
[100] Wohlwill, *op. cit.* (note 10), pp. 362-3.
[101] David Fabricius, *Prognosticon astrologicum 1615* (Nuremberg: Lauer), A2v-A3r.
[102] Ernst Zinner, "Zur Ehrenrettung des Simon Marius," *Vierteljahrsschrift der Astronomischen Gesellschaft*, 77 (1942), 23-75, p. 24.

January 1610,[103] but the year of printing is only given on the 1611 edition (1611, in fact, and the 1612 and 1613 prognostics also bear the years 1612 and 1613 as dates of printing).

It is certain, however, that, in the case of particular phenomena to be reported, also taking into account the needs of distribution of this kind of publication,[104] in the past, as in the present, news could be added up to a few months before the beginning of the year to which the prediction referred, and the dedication could also be changed at the last moment. All the more so if, as in Mayr's case, he had ready access to a publisher (in this instance, his father-in-law).[105] So the absence of any mention of Jupiter's satellites in the earlier prognostics is inexplicable. Zinner justifies the omission by stating that the 1610 *Prognosticon* was compiled in 1608 and that the 1611 *Prognosticon* was prepared in May 1609, but he offers no evidence in support of this thesis.[106] According to Zinner, therefore, the first publication in which Mayr could speak of the satellites was the 1612 *Prognosticon*, which was compiled (according to Zinner) in 1610. The evidence that Zinner brought to support this dating, however—the mention of a planetary configuration that occurred in that year—is puzzling.[107]

Among other things, along with the old prognostics, Zinner also found five letters from Mayr to Michael Maestlin in which Mayr discusses his own observations and celestial novelties.[108] In Mayr's letter of 6 December 1609, however, he made no mention of the telescope or of his first observations with it. The letter of 29 December 1611 mentions sunspots but not yet Jupiter. Jupiter makes its appearance only in the 29 March 1612 letter, in which Mayr reported:

> I have already completed the periods of the four planets of Jupiter observed by Galileo and myself for the first time and I have calculated the tables of their distances from both sides of Jupiter.

This, in fact, constitutes Mayr's first known claim of independent discovery, but it comes a full twenty-eight months after the first observations he mentioned in *Mundus Jovialis*.

**Evidence in Mayr's favour**

Mayr's supporters nonetheless claimed that he presented unobjectionable witnesses to his reliability in *Mundus Jovialis*. First of all, the dedication to the Margraves of Brandenburg is proof of his good faith. However, as we have seen, a fraudulent work, entirely or partially by Mayr, had already been dedicated to Margrave Joachim Ernest, Mayr's patron. It is also worth remembering that the Medici court itself, in whose honor Galileo had named Jupiter's satellites the "Medicea Sidera," had begun to doubt their existence once the initial enthusiasm following the publication of *Sidereus Nuncius* had died down. No one apart from the discoverer had confirmed seeing them, and the court asked Kepler, through his ambassador in Prague, Giuliano de' Medici,[109] for independent corroboration, which was delayed to the understandable concern of the Medici, who feared making a bad impression in front of the whole of Europe.[110] In short, the subject of a dedication obviously cannot be considered the guarantor of the content of the work any more than a poet's mistress is responsible for the quality of a poem dedicated to her!

---

[103] *Prognosticon astrologicum 1609* (Nuremberg: Lauer); *Prognosticon astrologicum 1610*; *Prognosticon astrologicum 1611* (Nuremberg: Lauer, 1611).
[104] See Jonathan Green, "The First Copernican Astrologer: Andrea Aurifaber's Practica for 1541," *Journal for the history of astronomy*, 41(2) (2010), 157-65, p. 157.
[105] Mayr had married Felicitas, the daughter of the Nuremberg printer and publisher Johann Lauer, who published most of his works, in 1606. See Hans Gaab, *op. cit.* (note 13), p. 106.
[106] Zinner, *op. cit.* (note 102), p. 34; in the sheets indicated in the Zinner table there is no useful reference, and, in the case of 1612, there are no sheets either.
[107] Ibid., p. 35.
[108] Ibid., pp. 40-45.

The second piece of evidence considered indisputable is the testimony of Mayr's protector, General Hans Philip Fuchs von Bimbach,[111] who helped him in his first attempts to obtain and work with the telescope. But the invocation of friends or public figures as witnesses, however prestigious, cannot support a scientific discovery. This is true today, but it was also true four centuries ago, as we have already seen with regard to sunspots. Kepler, in a manner that was imprudent to say the least, gave his support to Galileo in a letter of 19 April 1610,[112] though he had not directly observed what Galileo had discovered. In the meantime, however, all of Galileo's colleagues in Padua refused to look through the telescope,[113] and the professors in Bologna, gathered around Antonio Magini, whose reputation at the time was much greater than Galileo's, denied the validity of his discoveries.[114] In fact, a few months later, on 9 August, Kepler asked Galileo, probably a little repentantly, the names of the witnesses to his discoveries.[115] Even after Galileo responded that they were the Grand Duke and Giulio de' Medici, brother of the Ambassador in Prague,[116] Kepler was not satisfied. Nor was it sufficient that they were scholars of the highest calibre (Paolo Sarpi and Antonio Santini, for example)[117] because they were nonetheless friends of Galileo. In fact, satellites and other celestial novelties became reality, as is customary in scientific discoveries, only when Kepler and Clavius with the other mathematicians of the Collegio Romano, the highest scientific authorities of the time, were able to independently confirm their existence, in early September 1610[118] and April 1611 respectively.[119]

---

[109] See the dedication of *Johannes Kepleri Dissertatio cum Nuncio Sidereo* (Frankfurt: Palthenius, 1611), A2r-A2v. For a more detailed discussion of the matter, see Johannes Kepler, *Discussione col Nunzio Sidereo e Relazione sui quattro satelliti di Giove* (edited by Elio Pasoli and Giorgio Tabarroni) (Turin: Bottega d'Erasmo, 1972), pp. XXIV, XXV, 113.

[110] Vincenzo Giugni, the Medici palace cloakroom attendant, told Galileo on 5 June 1610 that the Grand Duke had instructed him to have a medal struck depicting the Medicean Planets, but to wait until the honour was justified by letters proving the discovery (in Galilei, *Opere*, 10, p. 368). This was in spite of the fact that, as Galileo replied rather angrily, the Grand Duke had "several times seen" Jupiter's satellites with his own eyes ("Galileo a [Vincenzo Giugni in Firenze]," 25 June 1610, in Galilei, *Opere*, pp. 379-82). In this letter, although Galileo formally praised his prince's prudence, he reported a long letter from Paris, written by an emissary of Henry IV of France on 20 April 1610 (the King died on 14 May), in which Henry, who was related to the House of Medici (he had married Maria de' Medici in 1600), asked Galileo to name the next celestial discovery after him!

[111] Mayr, *op. cit.* (note 2), *Praefatio*, 3r-3v.

[112] "Giovanni Kepler a [Galileo in Padova]," in Galilei, *Opere*, 10, pp. 319-40. This was later published by Kepler as *Johannes Kepleri Dissertatio cum Nuncio Sidereo* (see note 109).

[113] "Galileo a Giovanni Kepler [in Praga]," in Galilei, *Opere*, 10, p. 423.

[114] "[Martino Horki] a Giovanni Kepler in Praga" and "Giovanni Antonio Magini a Giovanni Kepler [in Praga]," in Galilei, *Opere*, X, pp. 343 and 359.

[115] "Giovanni Kepler a Galileo [in Padova]," in Galilei, *Opere*, 10, p. 416. 115. "Giovanni Kepler a Galileo [in Padova]," in Galilei, *Opere*, 10, p. 416.

[116] "Galileo a Giovanni Kepler [in Praga]," in Galilei, *Opere*, 10, p. 422.

[117] "[Paolo Sarpi a Giacomo Leschassier]," in Galilei, *Opere*, 10, p. 290; *Ioannis Antonii Roffeni epistola apologetica contra peregrinationem Martini Horki*, in Galilei, *Opere*, 3, p. 198.

**Technical impossibility**

The history of the development of the first telescopes and their technical capabilities is another factor that seriously undermines Mayr's credibility. The first instruments manufactured in the Netherlands in 1608 were little more than toys and were completely unsuitable for celestial observations. The same can be said for the telescopes that began to be manufactured in France and in various cities in Italy during 1609. On the other hand, the first telescopes built by Galileo were already on a par with these and soon surpassed them. We have an initial indication of the quality of Galileo's telescopes in the episode of an unnamed foreigner who visited Padua and Venice and who, in August 1609, wanted to sell one of the Dutch instruments to the Republic of Venice. The Venetian government gave Paolo Sarpi the task of investigating the potential purchase, which Sarpi ultimately counselled against. He knew that Galileo was trying to duplicate the Dutch instrument and had probably already seen that it would be equal to or better than those manufactured abroad.[120]

    Another clue to the quality of the telescopes of the time can be found in Harriot, who wrote in his notes that he turned a Dutch six-magnification telescope toward the Moon on the evening of 26 July 1609 (Julian date).[121] None of Galileo's lunar observations were dated, but it is likely that he began his observations in early September 1609, although the drawings later published in *Sidereus Nuncius* were made at least several weeks later.[122] Harriot's is therefore undoubtedly the first recorded telescope observation of any celestial body.[123] However, in Harriot's drawing of the Moon, we see less of Earth's satellite than can be discerned with the naked eye. Harriot's friend William Lower observed the Pleiades and the area of Orion's sword at the same time as Galileo (i.e., in the winter of 1609-1610), but he

---

[118] Informal mentions are found in several letters to Galileo, the first of which should be that of 6 September 1610 ("Giuliano de' Medici a Galileo in Firenze," in Galilei, *Opere*, 10, pp. 427-8); official recognition came with the publication, probably in November 1610, although it bears the date 1611, of *Ioannis Kepleri Narratio de observatis a se quatuor Jovis satellitibus erronibus* (Frankfurt: Palthenius, 1611)

[119] This was informally reported in several letters: 4 December 1610 ("Antonio Santini a Galileo in Firenze"), 17 December 1610 ("Cristoforo Clavio a Galileo in Firenze"), 22 January 1611 ("Cristoforo Grienberger a Galileo in Firenze"), in Galilei, *Opere*, 10 and 11, pp. 479-80, 484-5, 31-5; formal attestation 24 April 1611 ("I matematici del Collegio Romano a Roberto Bellarmino in Roma"), in Galilei, *Opere*, 11, pp. 92-3.

[120] *Opere del P.M.F. Paolo Sarpi* (Helmstedt: Müller, without date, prob. 1762), 1, p. 55; "Lorenzo Pignoria a Paolo Gualdo [in Roma], 1 August"; "Giovanni Bartoli [a Belisario Vinta in Firenze], 22 August"; "Giovanni Bartoli a Belisario Vinta in Firenze, 29 August," in Galilei, *Opere*, 10, pp. 250 and 255.

[121] Harriot, *op. cit.* (note 63), Vol. IX: The Moon [HMC 241 IX f. 26].

[122] Alberto Righini, "The Telescope in the making," in *Galileo's Medicean Moons: their impact on 400 years of discovery* (C. Barbieri, S. Chakrabarti, M. Coradini & M. Lazzarin, eds.), Proceedings IAU Symposium No. 269, 2010, 27-32.

[123] The famous bulletin printed in The Hague in October 1608 on the occasion of the appearance of the telescope included the statement that "…even the stars that do not ordinarily appear to our sight and eyes because of their smallness and the weakness of our sight, can be seen by means of this instrument" (*Ambassades du Roy de Siam envoyé a l'Excellence du Prince Maurice, arrivé à la Haye le 10 Septemb. 1608* [The Hague: 1608], B2r); the indication is not only undated but also overly general and vague, however.

saw only seven stars in the Pleiades[124] while Galileo correctly drew thirty-six.[125] Harriot returned to observing the Moon with better telescopes in July 1610, but his drawings of the various phases of the Moon are technically far inferior to Galileo's.[126]

Bartholomäus Schröter wrote to Galileo on 8 July 1610 to complain that Prince Augustus of Anhalt, despite sparing no expense, had not been able to obtain lenses suitable for the construction of a sufficiently good telescope in his region and that those he had obtained from France or Belgium were no better.[127]

When Kepler wrote to Galileo in August 1610 asking him to identify witnesses to the latter's observations, Kepler complained that there were no telescopes in the whole of Prague good enough to show the celestial novelties.[128] He only succeeded with the telescope that Galileo had sent to the Archbishop of Cologne, who lent it to Kepler.[129]

In September 1610, the Jesuits in Rome, using a telescope built by Giovan Paolo Lembo, were able to see the troubled face of the Moon as well as stars invisible to the naked eye in the Pleiades, Orion, and elsewhere,[130] but not Jupiter's satellites.[131] They succeeded thanks to a new telescope built by Lembo and were even more successful with one sent by the aforementioned Antonio Santini.[132]

Daniello Antonini wrote to Galileo from Brussels on 9 April and 2 September 1611:[133]

In these parts one does not find glasses that grow more than about 5 times the line....

I have seen some of the most exquisite glasses that are made in these parts; but they are worth nothing compared to that of V.S. which I saw in Padua, because there is no one who multiplies the line more than 10 times.... I have seen those of his own first inventor... but they are all cheap.

Thomas Harriot was able to see Jupiter's satellites for the first time only on 17 October 1610, and all four only on 14 December.[134] Johannes Fabricius, in June 1611, could only perceive the satellites indistinctly with his telescopes.[135]

---

[124] Chapman, *op. cit.* (note 63), p. 103.
[125] Vanin, *op. cit.* (note 28), pp. 458-9.
[126] Harriot, *op. cit.* (note 63), Vol. IX: The Moon [HMC 241 IX ff. 19-20].
[127] Galilei, *Opere*, 10, pp. 393-7.
[128] Kepler, *op. cit.* (note 115), p. 414.
[129] Kepler, *op. cit.* (note 118), A3r.
[130] Grienberger, *op. cit.* (note 119), pp. 33-4.
[131] "Galileo a [Cristoforo Clavio in Roma]," 17 September 1610, in Galilei, *Opere*, 10, p. 431.
[132] Grienberger, *op. cit.* (note 119), p. 34; "Antonio Santini a Galileo in Firenze," 4 December 1610, in Galilei, *Opere*, 10, pp. 479-80. Santini, who had succeeded as early as May, was soon able to build good telescopes, probably of the same quality as Galileo's.
[133] In Galilei, *Opere*, 11, pp. 84 and 204. In the continuation of the first letter, Antonini reported, "...however, in the past few days I had some irons worked for me, and after much effort I succeeded in making a pair of glasses that carries more than three and a half arms of a cannon, and with a mediocre concave the line grows about 40 times, and makes it very clear: so that I was able to observe very well the Medicean Planets and the unevenness of the Moon." He, like Santini, was one of the first to build telescopes of the same power as and probably of no lesser quality than those used by Galileo.
[134] Harriot, *op. cit.* (note 63), Vol. IV: Jupiter's satellites [HMC 241 IV ff. 3-4].
[135] Fabricius, *op. cit.* (note 62), B4r-B4v.

Van Helden reported[136] that Jacob Christmann saw Jupiter with his telescope at the beginning of December 1611 from Heidelberg and described it as "divided into three or four fiery balls, from which descended luminous wisps similar to comet tails."[137]

Gallanzone Gallanzoni wrote to Galileo from the Ardéche, in southern France, on 18 August 1612, commenting that there were no "good glasses" there,[138] and Michael Maestlin, in November 1614, had not yet succeeded in building a telescope that would show the satellites or phases of Venus.[139]

On the other hand, not even the first telescopes built by Galileo before the end of 1609 were able to show Jupiter's satellites, as he himself confirmed.[140] Even later, he was only satisfied with a minority of those he had built:

> … because the most exquisite glasses, capable of showing all observations, are very rare, and I, among more than 60 made at great expense and effort, have not been able to choose but a very small number of them....[141]

From the available evidence, then, it seems clear that Harriot was the only scientist north of the Alps with a telescope capable of showing Jupiter's satellites clearly, and his reports came almost a year after Galilei's first observations. No one in the Netherlands, France, Belgium, or Germany were able to achieve the same success at that time or for some time to come. In Prague, Kepler only succeeded thanks to a Galileo telescope. In Italy, only Galileo's or Santini's telescopes made observation of Jupiter's satellites possible. Consequently, it is hardly credible that Mayr had a telescope from "Belgium" capable of showing Jupiter's satellites in the summer of 1609 or that, toward the end of that year, an even better one was sent to him from Venice by Johann Baptist Lenck,[142] the Margraves' private advisor. No other support of this hypothesis exists, however, and all historical evidence points instead in a completely different direction.

At the same time, if Mayr had really discovered Jupiter's satellites independently at the end of 1609, why had he not immediately informed his friends and correspondents, Odontius, Vicke, Fabricius, Maestlin, and Kepler? Why had he not immediately spread the news through the press, given the availability of a publisher at home? If he had really made such a discovery, enormous outcry ought to have arisen in Germany and throughout Europe, as was the case with the publication of *Sidereus Nuncius.* Why would he have kept the news secret at first and then run the risk of not appearing credible by divulging it years later?

---

[136] Albert Van Helden, "The telescope in the seventeenth century." *Isis*, 65 (1974), pp. 38-58, 52.
[137] Jacob Christmann, *Nodus Gordius ex doctrina sinuum explicatus* (Heidelberg: Voegelin, 1612), p. 42.
[138] Galilei, *Opere*, 11, p. 378.
[139] Drake, *op. cit.* (note 10), p. 318.
[140] Galilei, *op. cit.* (note 67), p. 17r; see also Vasco Ronchi, "Sopra i cannocchiali di Galileo," *L'Universo*, 9 (1923), 791-804; *idem*, "Sopra le caratteristiche dei cannocchiali 'di Galileo' e sulla loro autenticità," *Rendiconti Reale Accademia Nazionale dei Lincei, Classe di Scienze Fisiche*, 32(2) (1923), 339-343.
[141] There are two versions of this letter, both autograph. The first states that "…I still have 10 glasses, which only, among the hundred and more that I have made with great expense and effort, are suitable to detect the observations in the new planets and fixed stars..." Galilei, *Opere*, 10, pp. 297-302.
[142] Mayr, *op. cit.* (note 2), *Praefatio*, 2v-3r.

## Satellite periods, Mayr's observations and tables

It is widely believed that the synodic periods of the satellites found by Mayr are better than those of Galileo[143] and that this would therefore confirm the superiority of his observations. This is not really the case. In Table 1, I show comparisons between modern values and the periods given by Galileo in his *Discorso delle cose che stanno in su l'acqua*,[144] those given by Mayr in *Prognosticon 1613*,[145] those reported in *Mundus Jovialis*,[146] and those deducible from the manuscripts of the calculations and tables of Galileo, who refined the values up to 1617.[147] From these figures it can be seen that, for two values, Mayr was closer than Galileo, and two others it is vice versa. This is probably the result of randomness (i.e., it is within the margins of error achievable with primitive instruments such as the first telescopes, which lacked the capability to make effective micrometric measurements).

| Satellite | Galileo 1612 | Mayr 1613 | Mayr 1614 | Galileo 1617 | Modern |
|---|---|---|---|---|---|
| I | 1d 18h and almost half | 1d 18h 18m 30s | 1d 18h 28m 30s | 1d 18h 28m 37s | 1d 18h 28m 36s |
| II | 3d 13h and about ⅓ | 3d 13h 18m | 3d 13h 18m | 3d 13h 17m 41s | 3d 13h 17m 54s |
| III | 7d and about 4h | 7d 3h 57m | 7d 3h 56m 34s | 7d 3h 58m 13s | 7d 3h 59m 36s |
| IV | 16d and about 18h | 16d 18h 23m | 16d 18h 9m 15s | 16d 17h 58m 41s | 16d 18h 5m 7s |

**Table 1**

With regard to Mayr's observations, we have only six. In chronological order, the first is dated 29 December 1609/8 January 1610, which presents some problems and was offered without measurements. In fact, at 5 p.m., the time Mayr says he made the observation,[148] the Sun was 3.8° below the horizon and the sky was too bright to see the satellites with a Galilean telescope.[149] The second is the one quoted in the letter to Odontius on the occasion of the lunar eclipse of 20/30 December 1610,[150] with quantitative measurements of the distance of the satellites expressed in absolute angular values, minutes and seconds. They are quite far from the true values because Mayr assigned to Jupiter an average apparent diameter of one minute, as previously mentioned.[151] If we divide the absolute values by the true apparent diameter of Jupiter, thus obtaining the relative values in units of Jupiter's diameter, however, we obtain good measurements whose errors are contained within 10%.

---

[143] Oudemans and Bosscha, *op. cit.* (note 25), p. 156; Rudolf Wolf, *Handbuch der Astronomie* (Zurich: Schulthess, 1892), 3, p. 464; Müller, *op. cit.* (note 10), p. 230; Anton Pannekoek, *A history of astronomy* (New York: Barnes and Noble, 1916), p. 231.
[144] Galilei, *op. cit.* (note 60), p. 63.
[145] Mayr, *op. cit.* (note 24), A4r. Presumably a printing error occurred in the first period: 18m instead of 28m.
[146] Mayr, *op. cit.* (note 2), A2r-A3r, Ev.
[147] Galileo Galilei, *Opere*. 2nd ed. edited by Antonio Garbasso (Florence: Barbera, 1929-39), 3.2, p. 898.
[148] Mayr, *op. cit.* (note 2), B4r.
[149] See Yaakov Zik, Giora Hon, and Ilan Manulis, "Did Simon Marius observe Jupiter's satellites on January 8, 1610? An exercise in computation," arXiv: 2002.04643. We have indirect evidence of this also in the letter by Galileo quoted above in note 131, which states that Jupiter could be seen through the telescope in broad daylight, with the Sun at a height of more than 15°, in the morning at the beginning of September 1610, but that the satellites could be seen only until the naked eye could make out Sirius, which has a magnitude of -1.4. Instead, with the Sun at -3.8°, the stellar magnitude limit is around -3.2, which means that only Venus can be seen, if it is present in the sky. Certainly this does not prove that Mayr could not have seen the satellites on that date, because ten or fifteen minutes later the sky would have been dark enough to allow the observation, provided one had a good enough telescope, but it shows his unreliability in reporting the time of an observation

Four other observations appeared in *Mundus Jovialis*, all from 1613. The first one, of 7-17 February,[152] lacks quantitative measurements. Mayr used the other three, of 20-30 January, 14-24 February, and 1-11 April,[153] as an example for the reader on the use of his tables, but these observations all contain quantitative measures of distance expressed in absolute angular values that are rather far from the actual values. If these figures are reduced by the value assumed from time to time by Jupiter's diameter, better data result with errors almost always contained within 10%. However, the time indication of one of these observations is also wrong—this time by quite a lot. On February 14-24 at 7 p.m., in fact, Jupiter was 3° below the horizon in Ansbach!

The fact that neither *Mundus Jovialis* nor any other of Mayr's publications show graphical representations of the satellites similar to those depicted by Galileo remains, however, quite strange, as others have pointed out.[154] Four observations with measurements alone, even if they can be considered equivalent to Galileo's in terms of precision, hardly constitute a representative sample, especially if compared to the sixty-five observations Galileo published in *Sidereus Nuncius* and to the more than 1,200 in his notes.[155]

Finally, regarding the precision of the tables, Cassini wrote that "the configurations taken from these tables bore no resemblance to the real configurations."[156] Oudemans, limiting his examination to the years 1609 to 1614, noted that errors remained on average below the arcminute.[157] Klug wrote that precision decreased from 1610 to 1614, taking this as evidence of Mayr's dependence of Galileo's tables and not on his own observations.[158] Drake stated that Mayr's tables were "quite good."[159]

I have completed about forty cross-checks for the period from 1609 to 1630 using *Starry Night*, *Guide*, and Meeus's tables, and the results are much worse than the examples, based on observations, reported by Mayr himself in *Mundus Jovialis*. As mentioned, for some obscure reason Mayr decided to give the distances in terms of absolute values, and the predictions he obtained were very far from the truth. Even if we reduce the data in terms of relative values, however, the accuracy is poor, and only in a third of the cases is the error

---

that was of such great importance for him. Obviously, he should have observed at least at the end of the astronomical twilight, 18:34 at Ansbach, with a dark sky and the planet much higher on the horizon.

[150] See note 97.

[151] All data on observations and tables were cross-checked with *Starry Night Pro Plus 8*, *Guide 9.1* and Jean Meeus's tables ("Tables of the satellites of Jupiter," *Journal of the British Astronomical Association*, 72, 1962, pp. 80-88). The differences in distance values obtained with the three modes are at most a few arcseconds.

[152] Mayr, *op. cit.* (note 2), D4v.

[153] Ibid., F1rv-F2rv.

[154] See, e.g. Drake, *op. cit.* (note 86) p. 92.

[155] No notes by Mayr have ever been found.

[156] *Diverse ouvrages d'astronomie par M. Cassini* (The Hague: Gosse and Neaulme, 1731), p. 412.

[157] Oudemans and Bosscha, *op. cit.* (note 25) pp. 168-72.

[158] Klug, *op. cit.* (note 17), pp. 498-508.

[159] Drake, *op. cit.* (note 86), p. 91.

contained within about 20%, without appreciable differences over the years. Because Mayr's tables appeared to me to be completely useless, I did not consider it appropriate to proceed with further checks.

**Conclusions**

In conclusion I think I can state:

1) Mayr attempted to claim, albeit in a somewhat convoluted manner, the first discovery of Jupiter's satellites.
2) There is no evidence that Mayr independently discovered the satellites; indeed a great deal of evidence exists to the contrary, first among which are the words Mayr himself provided in his writings.
3) There is no evidence that Mayr observed the satellites before 30 December 1610, the date mentioned in his letter to Odontius. Indeed, numerous evidence exists to the contrary, including Mayr's own writings, Galileo's survey in *Il Saggiatore* on the position of the satellites in the first two years, and the technical level of the telescopes of Mayr's time.
4) Mayr probably did not observe Jupiter with accuracy and continuity before 1613, and, in any case, not enough to ensure a wealth of observations to establish tables of sufficient precision.[160]
5) There are suspicions that Mayr based his discoveries on the "World of Jupiter" on Galileo's observations rather than on his own.

Mayr's supporters nonetheless praise him for his tables and for his foresight and for his ability to achieve a result in line with the technological possibilities of the time.[161] In my view, conversely, Mayr was slightly reckless in publishing tables that could not be useful because they were based on observations made with instruments that were inadequate for the task (anyone who has tried to observe Jupiter's satellites with a Galilean telescope of a 20-30 magnification knows this). Galileo, on the other hand, after recording more than a thousand observations between 1610 and 1619, realised that it would never be possible to arrive at satisfactory tables and gave up in the face of the titanic task.

Mayr's followers also generally think he was unlucky in the mistreatment he received at the hands of his contemporaries and of history.[162] This may be so, but it may be equally true that his memory has been preserved beyond his merits: an asteroid has been dedicated to him, a long rille on the Moon and a crater seven times larger than the one named after Galileo[163] bear Mayr's name, and the names he gave to satellites will remain in history

---

[160] Drake, the only one who has dealt with this issue in depth in modern times, stated (*op. cit.* in note 86, p. 92) that at least three years of continuous observation would have been required to succeed in the task.
[161] See, e.g., Bosscha, *op. cit.* (note 7), p. 518.
[162] See, e.g., Oudemans and Bosscha, *op. cit.* (note 25), p. 167; Zinner, *op. cit.* (note 102), p. 40; Jay M. Pasachoff, "Simon Marius's *Mundus Jovialis*: 400th anniversary in Galileo's shadow," *Journal for the history of astronomy*, 46(2) (2015), 218-34, pp. 227 and 231.
[163] Also the topography of this lunar region has remained shrouded in historiographical confusion. In their map, Grimaldi and Riccioli assigned Galileo a crater 3.5 times larger than the one dedicated to Mayr: see Giovanni Battista Riccioli, *Almagestum novum* (Bologna: Benati 1651), 1, plates outside the text between pp. 205 and 205½. Tobias Mayer, in his map, reduced the crater dedicated to Mayr, and assigned to Galileo what Grimaldi

forever.[164] They are beautiful, entirely appropriate names, and it would probably not have been possible to think of better ones. I think that is enough.

## Acknowledgements

The author would like to thank Prof. Elena Avanzo and Prof. Emma Dee Steed for revision of the text.

---

and Riccioli had dedicated to Reinieri; see Ewen A. Whitaker, *Mapping and Naming the Moon* (Cambridge: Cambridge University Press, 1999), pp. 84, 102-3. Beer and Madler later discovered that no crater exists in the area indicated by Grimaldi and Riccioli, but only a zone with a higher albedo. Beer and Madler then assigned Galilei the largest crater in the region, which was rather small: see Wilhelm Beer and Johann Heinrich Madler, *Der Mond* (Berlin: Schropp, 1837), p. 282.

[164] Cassini, too, in the *Ephemerides bononienses mediceorum siderum* (Bologna: Manolessi, 1668), p. 11, proposed mythological names, equally or perhaps more appropriate because they were inspired by Jupiter's more "institutional" loves and did not evoke acts of deception, kidnapping, or rape like the names that have been adopted: Pallas (epithet of Minerva, daughter of Jupiter), Juno (Jupiter's third wife), Themis (Jupiter's second wife), Ceres (who joined with Jupiter to generate Proserpine).